%% file: main.tex
\documentclass[a4paper,oneside, preprint,12pt]{elsarticle}

\input{Config/packages}

\input{Config/macros}


\journal{Scientific Reports}

\begin{document}

\begin{frontmatter}

\title{An Innovative Photon-Driven Subcritical Reactor Concept Powered by Synchrotron Radiation Source } %

\author[First,Second]{\texorpdfstring{Antonio Cammi\corref{cor1}}{Antonio Cammi}}
\author[First]{Lorenzo Loi}
\author[Third,First]{Andrea Missaglia}
\author[Fourth]{David Alesini}
\author[Fifth]{Hans-Heinrich Braun}

\cortext[cor1]{Corresponding author. Email address: antonio.cammi@polimi.it}
\address[First]{Politecnico di Milano, Department of Energy, Via La Masa 34, Milano 20133, Italy}
\address[Second]{Emirates Nuclear Technology Center (ENTC), Department of Mechanical and Nuclear Engineering, Khalifa University, Abu Dhabi, 127788, United Arab Emirates}
\address[Third]{Department of Nuclear Engineering, Texas A\&M University, 423 Spence St, College Station, 77843-3133,  TX, USA}
\address[Fourth]{INFN–Laboratori Nazionali di Frascati, Via E. Fermi 54, Frascati (Roma) 00044, Italy}
\address[Fifth]{Paul Scherrer Institut, Forschungsstrasse 111 5232 Villigen PSI, Switzerland}

\begin{abstract} 

Most reactors are fuelled with enriched uranium, whose production is proliferation-sensitive and concentrated in a few suppliers, and discharge fuel that still contains fissile material and accumulates in storage. Accelerator-driven subcritical reactors relax the requirements on the fuel, but proton beams must travel in vacuum to a spallation target inside the core. Here we show that synchrotron radiation can drive a subcritical heavy-water reactor fuelled with natural uranium or with spent fuel loaded without reprocessing. A photon beam with a critical energy of 10~MeV, from a proposed 32~GeV electron storage ring, releases neutrons in a uranium converter at the centre of the core. Coupled photon–neutron Monte Carlo simulations show that the two cores, loading different fuels but being at a multiplication factor close to 0.985, reach the same gain: one 400~kW beamline drives each of them close to 4.4 MW of fission power, about 11 times the beam power. A storage ring with 50 beamlines could drive one core from each beamline, every core controlled through its own photon beam. These results make synchrotron radiation a candidate driver for thermal subcritical reactors that need neither enrichment nor reprocessing, complementary to proton-driven fast systems for transmutation.

\end{abstract}

\end{frontmatter}


\input{Sections/01_introduction}

\input{Sections/02_results}
\input{Sections/03_discussion}

\input{Sections/04_methods}


\section*{Data availability}
The OpenMC input files of every simulation, the photon source spectrum, the results of each design in machine-readable form (JSON) and the data underlying every figure and table are deposited in the GitHub repository \url{https://github.com/ERMETE-Lab/synchrotron-driven-reactor}, which will be made publicly available upon publication. The OpenMC statepoint files are available from the corresponding author on reasonable request. The nuclear data used are publicly available: ENDF/B-VII.1~\cite{ENDFVII1} for neutron, photon and photonuclear interactions, and the IAEA Photonuclear Data Library 2019~\cite{Kawano_2020}.

\section*{Code availability}
The model and post-processing code (Python, OpenMC Python API) and the Jupyter notebooks that produce every result, figure and table of this article are deposited at \url{https://github.com/ERMETE-Lab/synchrotron-driven-reactor} under the MIT licence and will be made publicly available upon publication.


\section*{Author contributions}
A.C. and L.L. conceived the concept. A.C. supervised the work. L.L. developed the methodology and the Monte Carlo
model, performed the simulations and the analysis, and wrote the manuscript. A.M. contributed to the concept. D.A.
and H.-H.B. suggested the electron storage ring concept and provided estimates of the ring and photon source
parameters. All authors reviewed and edited the manuscript.

\section*{Competing interests}
The authors declare no competing interests.


\clearpage
\section*{Supplementary Information}
\setcounter{table}{0}
\renewcommand{\thetable}{S\arabic{table}}
\input{Tables/photoneutrons}    
\input{Tables/sf_composition}   

\newpage
\bibliography{bibliography}

\end{document}

%% file: Config/packages.tex
\usepackage{url}
\usepackage[nopatch=footnote]{microtype}

\usepackage[a4paper, margin=2.5cm]{geometry}
\usepackage[utf8]{inputenc}
\usepackage{natbib}
\usepackage{amsmath}
\usepackage{amsthm}
\usepackage{amssymb}
\usepackage{amsfonts}
\usepackage{bm}
\usepackage{siunitx}
\usepackage{epstopdf}
\usepackage{flushend}
\usepackage{parskip}
\usepackage[hyperfootnotes=true,colorlinks=true,linkcolor=blue, bookmarks=true,citecolor={red},urlcolor=cyan]{hyperref}

\usepackage[pagewise]{lineno}
\usepackage{framed} 
\usepackage{multicol} 
\usepackage{tabularx}
\usepackage{longtable} 
\usepackage[ruled,vlined]{algorithm2e}
\usepackage{multirow}

\usepackage{amssymb}

\usepackage[flushright]{rotating}

\usepackage{subcaption} 

\usepackage[printonlyused]{acronym} 
\usepackage{mathtools} 

\usepackage[dvipsnames]{xcolor}
\usepackage{tikz}
\usetikzlibrary{fit}
\usetikzlibrary{positioning}
\usetikzlibrary{shapes,arrows}
\usetikzlibrary{backgrounds}
\usetikzlibrary{calc}
\usetikzlibrary{arrows.meta}

\usepackage{setspace} 
\usepackage{emptypage} 

\usepackage{booktabs}
\usepackage{comment}

%% file: Config/macros.tex
\definecolor{bluePolimi}{RGB}{22, 44, 80}
\definecolor{lightBluePolimi}{RGB}{91, 122, 172}
\definecolor{greenPolimi}{RGB}{0, 110, 0}
\definecolor{redPolimi}{RGB}{180, 0, 0}

\newcolumntype{M}[1]{>{\centering\arraybackslash}m{#1}}

%% file: Sections/01_introduction.tex
\section{Introduction}\label{sec:intro}

Nuclear fission supplies about 9\% of the world's electricity from more than 400 power reactors, and its role is set to grow in the coming decades: installed capacity is projected to increase by 50--160\% by 2050, and 31 countries have pledged to triple it by the same date~\cite{IAEA_RDS1_2025}. Most operating reactors are light-water reactors fuelled with uranium enriched from the natural 0.7\% to 3--5\% in $^{235}$U~\cite{WNA_enrichment}. As critical systems, they discharge their fuel when its residual reactivity can no longer sustain the chain reaction or its burnup reaches the limits set by the integrity of the cladding~\cite{WNA_fuel}, although it still contains fissile material. The expansion of nuclear power therefore depends on two stages of the fuel cycle: the \textit{enrichment of fresh fuel} and the \textit{management of spent fuel}.

Enrichment is one of the most sensitive steps of the front end of the fuel cycle for non-proliferation. Since the step from reactor grade to weapons grade is relatively small, enrichment plants are under tight international control, and four suppliers hold more than 99\% of the world's capacity, a concentration that represents a risk for the expansion of nuclear power~\cite{IEA_Nuclear2025, WNA_enrichment}. A reactor fuelled with natural uranium would remove this step from its fuel cycle altogether.

At the back end of the cycle, spent fuel accumulates without a final destination. About 448000~t has been discharged from commercial power reactors and less than a third of it has been reprocessed~\cite{WNA_processing}; the rest is held in interim storage, since no geological repository is yet in operation~\cite{Posiva_2026}.
Yet this material is not exhausted as a fuel: besides fission products, it still contains part of its $^{235}$U and up to 1\% of plutonium~\cite{WNA_processing}. Today it can be used again only after reprocessing, performed on an industrial scale in only a few countries~\cite{WNA_processing}. Reactors moderated by heavy water, which need less fissile material, offer a more direct route: spent fuel has been converted into fuel for these reactors by mechanical and thermal treatment alone, without chemical separation~\cite{Yang_2006}. A reactor able to use spent fuel without reprocessing would turn part of this inventory into a fuel resource.

The need for enrichment and the discharge of fuel that still contains fissile material have a common origin. In a critical reactor the fuel must sustain the chain reaction on its own, which confines its fissile content to an operating window: at loading it must be high enough to provide the excess reactivity for the whole irradiation, which in a light-water lattice requires enrichment, and once burnup has consumed this excess the fuel must be replaced, although it still contains fissile material. Subcritical operation removes this constraint. In an accelerator-driven system (ADS), the core is kept below criticality and the neutrons it lacks are supplied by an external source~\cite{Nifenecker_2001}. Without the source the fission chains die out: the power follows the driver and drops promptly to the decay heat when the driver stops, while the margin below criticality protects the core from reactivity insertions that would drive a critical reactor into a power excursion~\cite{Carminati_1993, Roche_1995}. The reactor is thus controlled through its source rather than through the reactivity of the core~\cite{Roche_1995}. The same margin relaxes the requirements on the fuel, and subcritical cores have been studied with thorium~\cite{Carminati_1993, Lin_2013} and with the actinides of spent fuel~\cite{Rubbia_1997}. Pressurised heavy-water reactors burn natural uranium~\cite{IAEA_TRS407_2002} but, as critical systems, need on-power refuelling and active reactivity control; subcritical operation moves this complexity to the source and reduces the core to a simple lattice of fuel, coolant and moderator.

Most ADS designs use a spallation source driven by a proton accelerator, from the thermal-spectrum system of Bowman et al.~\cite{Bowman_1992} to the thorium-fuelled Energy Amplifier of Rubbia and co-workers~\cite{Carminati_1993, Roche_1995, Rubbia_1997}. The durability of the spallation target and the reliability of the accelerator have been the main difficulties of these designs~\cite{Kadi_2006, Marques_2024}, and the projects now under way, MYRRHA~\cite{Baeten_2014}, CiADS~\cite{deng2025preliminary} and LEHIPA~\cite{priyadarshini2024phase}, are research facilities. Photonuclear neutron sources are an alternative that is attracting growing interest. In electron-driven systems, the neutrons are released by the photonuclear reactions of the bremsstrahlung photons that the electrons emit in a high-$Z$ converter~\cite{Petwal_2007}: a subcritical assembly of 19.7\% enriched uranium in light water, driven by a 100~MeV, 100~kW electron linac, has been built at the Kharkiv Institute of Physics and Technology~\cite{Gohar_2022, Karnaukhov_2024}, and studies have considered molten-salt cores~\cite{Lin_2013, Rummana_2024}, heavy-water systems in which deuterium also acts as a photoneutron converter~\cite{Feizi_2016}, and subcritical versions of PWR and small modular cores~\cite{Liu_2022, Firoozabadi_2022}. More recently, gamma-ray beams from laser-Compton scattering and from the Gamma Factory proposed at CERN have been studied as drivers of cores of uranium enriched to 23--25\%, for the transmutation of long-lived fission products; they require about $10^{19}$~photons~s$^{-1}$, about two orders of magnitude beyond present gamma-ray sources~\cite{Sun_2022, Hu_2025}.

Here we show that synchrotron radiation can drive a subcritical thermal reactor fuelled with either natural uranium or spent fuel. The driver is the photon beam emitted by the high-field dipoles (super-bends, S-BENDs) of a proposed high-energy electron storage ring whose radiation has a critical energy of 10~MeV, a photonuclear neutron source whose yield has been characterised in isolated targets~\cite{SYNERGY_2026}. The photons are generated in the ring and transported to the reactor along a beamline, so that, unlike in electron-driven systems, neither the electron beam nor the bremsstrahlung converter that stops it enters the core. The complexity is concentrated in the accelerator, and the power of each core follows the intensity of its own photon beam, which can be stopped upstream of the core. We couple a single 400~kW photon-beamline, to a lattice of oxide fuel pins cooled and moderated by heavy water, without control rods, at $k_\text{eff}=0.985$, and study two versions of the core that differ only in the fuel and in the lattice pitch: natural UO$_2$, which requires no enrichment, and PWR spent fuel discharged at 30~GWd~t$^{-1}$ and cooled for 5~years~\cite{Takano_1994}, loaded without reprocessing. In both versions, the beam stops in a natural-uranium converter at the centre of the core, where photonuclear reactions release the source neutrons. With coupled photon--neutron Monte Carlo simulations, we show that the two versions amplify the beam power by almost the same factor, although the tighter spent-fuel lattice has a harder neutron spectrum and a more peaked power distribution. The same concept can therefore run on natural uranium without enrichment or on spent fuel without reprocessing, and one storage ring could drive several such cores.

%% file: Sections/02_results.tex
\section{Results}\label{sec:results}

\subsection{Photon-driven subcritical core}\label{sec:res-design}

The performance of the system is quantified by the power gain
\begin{equation}
\label{eqn:gain}
    G = \frac{P_{core}}{P_{beam}},
\end{equation}
the ratio between the thermal power $P_{core}$ released by fissions in the core and the photon beam power ($P_{beam}$) that drives it, 400~kW, in the analized case and emitted by one S-BEND. It determines the net energy balance of the facility and allows a direct comparison with subcritical systems driven by spallation or electron beam bremsstrahlung sources (Section~\ref{sec:discussion}).

\begin{figure}
    \centering
    \includegraphics[width=0.9\linewidth]{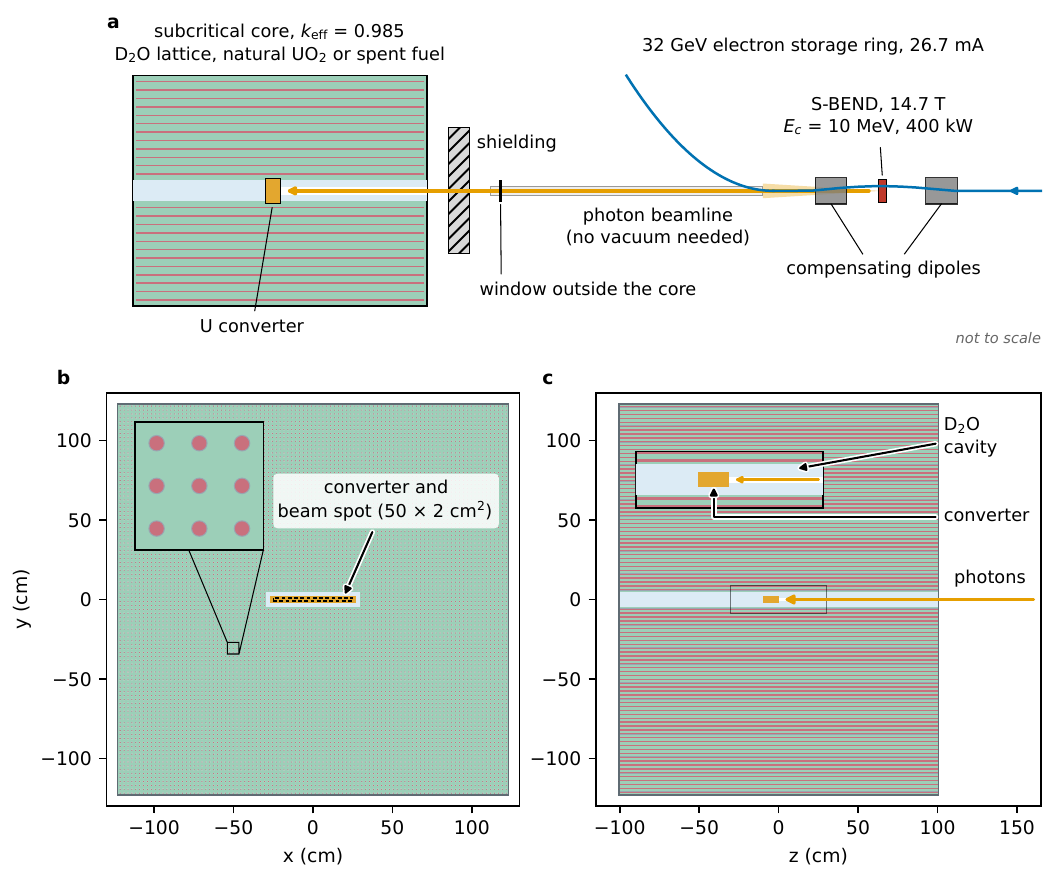}
    \caption{Photon source, beamline and core. \textbf{(a)} Concept, not to scale. The electrons of a 32~GeV storage ring with a stored current of 26.7~mA cross one of its 50 S-BEND modules, a 14.7~T dipole between two compensating dipoles of opposite polarity that close the orbit bump. The S-BEND emits a 400~kW photon beam with a critical energy of 10~MeV, which travels in air through a beamline closed by a window outside the core and its shielding, and stops in the uranium converter at the centre of the core. \textbf{(b, c)} Cross sections of the spent-fuel core from the Monte Carlo model; the natural-uranium core has the same layout with a lattice pitch of 2.835~cm instead of 2.450~cm. \textbf{(b)} $x$--$y$ plane perpendicular to the beam, through the converter; the dashed rectangle marks the $50\times2$~cm$^2$ beam spot, and the inset enlarges a $3\times3$ block of pins. \textbf{(c)} $z$--$y$ plane through the pin row next to the core centre, with the converter region enlarged in the inset. The photon beam enters from the right ($z=100$~cm) through a channel, void in the model, and stops in the natural-uranium converter ($-10<z<0$~cm). Rose, oxide fuel pins; sage green, heavy-water coolant and moderator; pale blue, heavy water in the central cavity; amber, converter; dark grey, 1~cm graphite layer.}
    \label{fig:geometry}
\end{figure}

The two designs, fuelled with natural UO$_2$ and with pressurised-water-reactor spent fuel, share the same geometry, photon beam and converter (Fig.~\ref{fig:geometry}): a horizontal square lattice of $100\times100$ Zircaloy-clad oxide pins, 200~cm long and parallel to the beam, cooled and moderated by heavy water. The beam enters a central cavity filled with heavy water through a channel and stops in a natural-uranium converter 54~cm wide and 10~cm thick, just beyond mid-length (Section~\ref{sec:m-model}). The lattice pitch is the only design variable and was adjusted in each case to obtain $k_\text{eff}=0.985$: 2.835~cm for natural uranium and 2.450~cm for spent fuel, whose higher fissile content (1.96~wt\% of $^{235}$U, $^{239}$Pu and $^{241}$Pu, against 0.71~wt\%) allows a tighter lattice. The two cores therefore contain nearly the same mass of heavy metal, 9.4 and 9.3~t, whereas the spent-fuel core is 245~cm wide instead of 284~cm and requires 11.6~t of heavy water instead of 16.1~t. The main results for both designs are summarised in Table~\ref{tab:key_results}.

\input{Tables/key_results}

\subsection{Power gain of the two designs}\label{sec:res-gain}

Driven by the same 400~kW beam, which corresponds to $S_0=3.76\times10^{17}$ photons~s$^{-1}$, the natural-uranium core produces $P_{core}=4.37\pm0.06$~MW and the spent-fuel core $4.41\pm0.07$~MW (Table~\ref{tab:key_results}), corresponding to $G=10.92\pm0.14$ and $11.03\pm0.16$, with the photonuclear data of IAEA/PD-2019 (Section~\ref{sec:m-model}). The two gains differ by $0.10\pm0.22$ and are therefore equal within the statistical uncertainty. Since the photon beam and the converter are identical, both cores are driven by the same external source, $Y_1=(6.65\pm0.02)\times10^{-3}$ first-generation photoneutrons per photon, or $S=S_0Y_1=2.50\times10^{15}$~n~s$^{-1}$ (see Section~\ref{sec:methods}), and the gain depends only on the response of each lattice to it. The fission and capture gammas of the core release further photoneutrons, mainly through D$(\gamma,n)$, which raise the steady-state yield $Y_n$ 11\% and 12\% above $Y_1$ (Table~\ref{tab:key_results}); being produced by the neutron population of the core, they belong to its multiplication rather than to the external source.

Both cores operate at $k_\text{eff}=0.985$, about 1500~pcm below criticality. A point-reactor estimate based on $k_\text{eff}$ alone, which gives the source neutrons the same importance as the fission neutrons ($\varphi^*=1$ in Eq.~\eqref{eqn:power}), would predict about half of the power computed for both cores. The difference comes from the position of the source: the source neutrons are released in the converter, at the centre of the core, where their importance is highest, and the source importance is $\varphi^*=1.90\pm0.04$ for natural uranium and $2.04\pm0.05$ for spent fuel (Eq.~\eqref{eqn:phistar}). Correspondingly, the source multiplication factor, $k_s=0.99246\pm0.00010$ and $0.99301\pm0.00011$, is well above $k_\text{eff}$. Only the coupled photon--neutron transport, which follows the source neutrons from where they are released, resolves this effect. The two factors have different roles: $k_\text{eff}$ is a property of the lattice alone and sets the margin to criticality, and with it the safety of the core, whereas $k_s$ depends on the position and energy of the external source. Placing the source where the importance is highest therefore raises the power obtained from a given beam without bringing the core closer to criticality, so that a gain of about eleven is obtained without reducing this margin.

The two cores reach the same gain through compensating effects (Section~\ref{sec:m-gain}). In the spent-fuel core, the stronger absorption of plutonium and fission products confines the fission chains closer to the converter, which raises the source importance and, as shown below, the peaking of the power, but the larger number of neutrons per fission of plutonium, $\bar\nu=2.68$ against 2.44, means that the same neutron population is sustained by fewer fissions.

\subsection{Neutron energy spectrum}\label{sec:res-spectrum}

Heavy water serves two purposes in both designs. As a moderator, its large slowing-down power and very low absorption allow a lattice of natural uranium to approach criticality without enrichment. As a neutron source, it releases photoneutrons throughout the lattice from the photons above the 2.224~MeV threshold of D$(\gamma,n)$, both from the beam and from the core, which raise the multiplication factor by about 85~pcm, a contribution that a neutron-only criticality calculation would miss (Section~\ref{sec:m-feedback}).

\begin{figure}[htbp]
  \centering
  \includegraphics[width=0.8\textwidth]{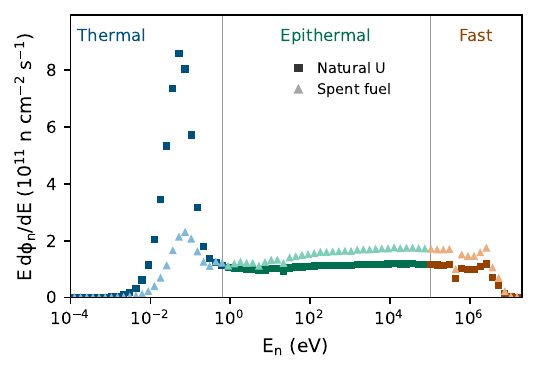}
  \caption{Neutron flux per unit lethargy, $E\,\mathrm{d}\phi_n/\mathrm{d}E$, averaged over the core lattice (fuel pins and heavy water), for natural uranium (dark squares) and spent fuel (light triangles), driven by one 400~kW S-BEND beamline. Colours and vertical lines mark the thermal ($E_n<0.625$~eV), epithermal and fast ($E_n>100$~keV) groups. One point every three energy bins (20 bins per decade); error bars, one standard deviation (statistical).}
  \label{fig:spectrum}
\end{figure}

The neutron spectra averaged over the core lattice (Fig.~\ref{fig:spectrum}) differ markedly between the two designs. The natural-uranium core is well thermalised: its spectrum shows a pronounced Maxwellian peak at a few tens of meV, and thermal neutrons account for about half of the flux. In the spent-fuel core the thermal peak is about four times lower and the thermal fraction drops to about one fifth, while the epithermal flux per unit lethargy is about 40\% higher. Two effects contribute to this hardening. First, the tighter lattice reduces the moderator-to-fuel volume ratio from 13.9 to 10.0, so that neutrons undergo fewer collisions with deuterium before being absorbed. Second, plutonium and the fission products introduce a strong absorption in the thermal and low-epithermal range, through the 0.3~eV resonance of $^{239}$Pu, the 1~eV resonance of $^{240}$Pu and the large thermal cross sections of $^{149}$Sm and $^{155}$Gd. Above 100~keV the two spectra have the same shape, but the spent-fuel spectrum is about 50\% higher: the spent-fuel core produces 8\% more fission neutrons in a lattice whose volume is 25\% smaller.

\subsection{Spatial distribution of the neutron flux}\label{sec:res-flux}

\begin{figure}[htbp]
  \centering
  \includegraphics[width=\textwidth]{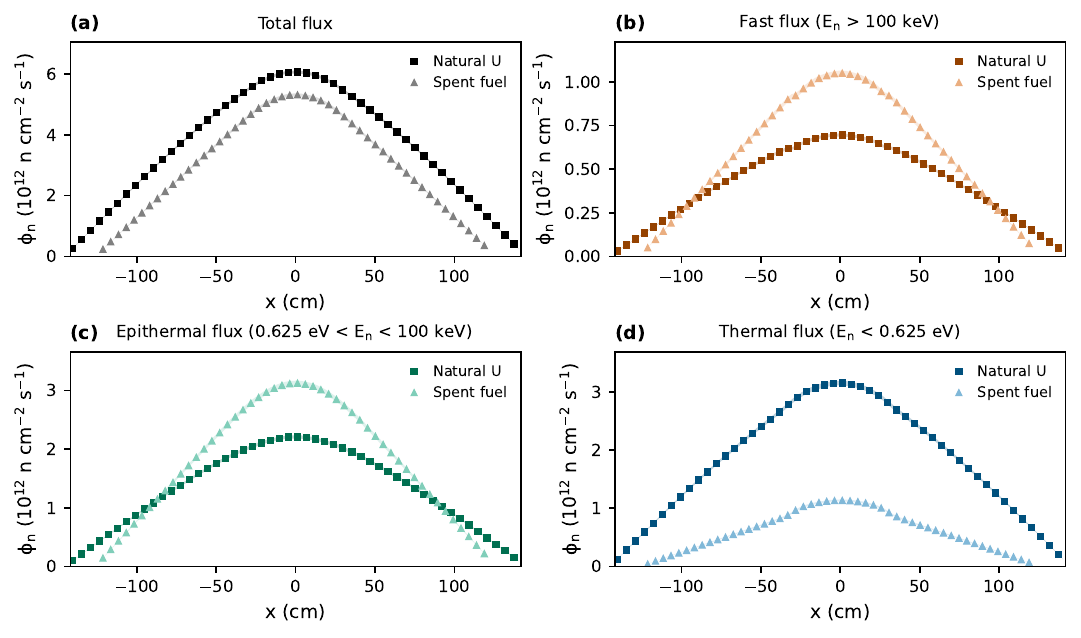}
  \caption{Neutron flux $\phi_n$ along $x$, averaged over $y$ and $z$ over the whole core, for natural uranium (dark squares) and spent fuel (light triangles), driven by one 400~kW S-BEND beamline. \textbf{a},~total flux; \textbf{b},~fast flux ($E_n>100$~keV); \textbf{c},~epithermal flux ($0.625$~eV$\,<E_n<100$~keV); \textbf{d},~thermal flux ($E_n<0.625$~eV). The converter sits at the centre ($|x|<27$~cm). The spent-fuel core is narrower because of its smaller lattice pitch. Error bars, one standard deviation (statistical), smaller than the markers.}
  \label{fig:flux-x}
\end{figure}

Figure~\ref{fig:flux-x} shows the flux profiles along $x$, averaged over $y$ and $z$. In both cores the flux is maximum at the centre, where the source neutrons enter the lattice, and decreases towards the boundary. The total flux at the centre is $6.1\times10^{12}$~n~cm$^{-2}$~s$^{-1}$ in natural uranium and $5.3\times10^{12}$~n~cm$^{-2}$~s$^{-1}$ in spent fuel, of which 52\% and 21\% are thermal. The fast and epithermal profiles of the spent-fuel core are higher at the centre but narrower, and fall below those of natural uranium near the edge of the smaller lattice.

\begin{figure}[htbp]
  \centering
  \includegraphics[width=0.9\textwidth]{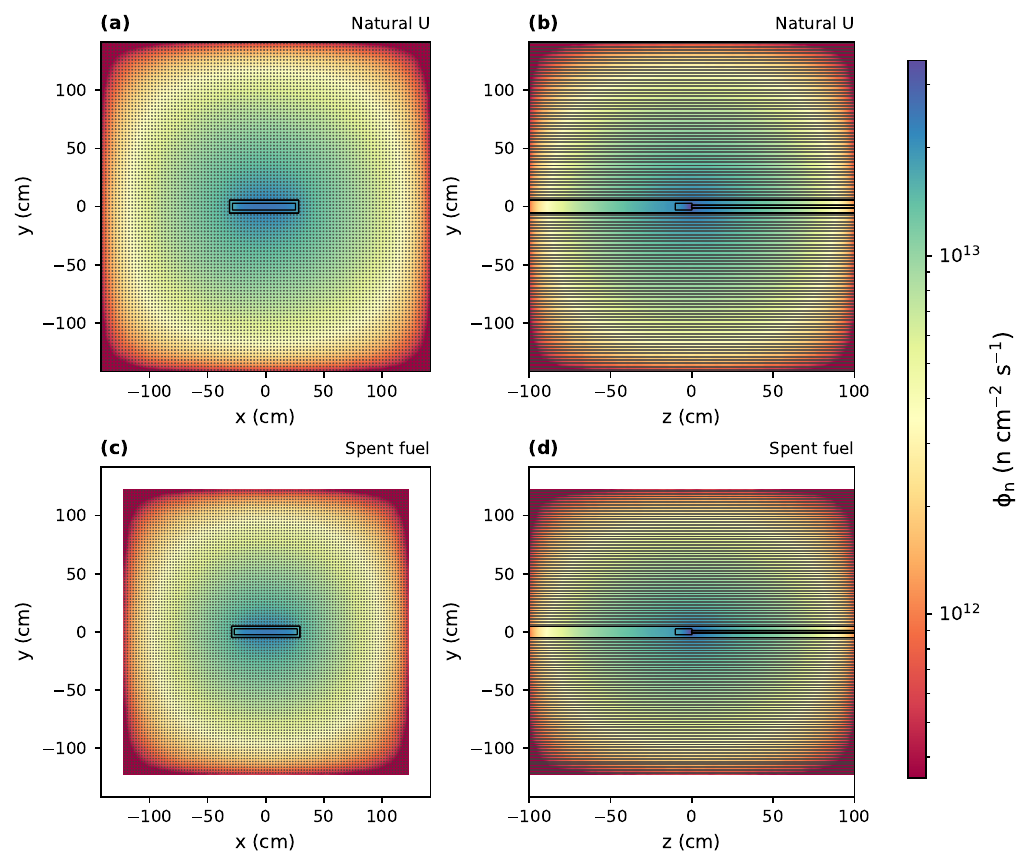}
  \caption{Total neutron flux per pin position, on a common logarithmic colour scale, for one 400~kW S-BEND beamline. \textbf{a},~\textbf{b},~natural uranium; \textbf{c},~\textbf{d},~spent fuel. \textbf{a},~\textbf{c},~$x$--$y$ plane, averaged over the length of the converter; \textbf{b},~\textbf{d},~$z$--$y$ plane through the centre of the core. Black lines: fuel pins (thin) and the converter, the cavity around it, the beam channel in front of it and the edge of the lattice (thick). The photon beam enters from the right ($z=100$~cm) and stops in the converter ($-10<z<0$~cm).}
  \label{fig:flux-maps}
\end{figure}

In the flux maps of Fig.~\ref{fig:flux-maps}, the flux peaks in the converter at about $3.5\times10^{13}$~n~cm$^{-2}$~s$^{-1}$ and decreases by about two orders of magnitude towards the boundary of the lattice. The localisation of the source in the converter produces a flux distribution more peaked than the fundamental mode of the core, which underlies the source importance above unity found in both cases. On a common colour scale the two distributions have the same shape, the spent-fuel core being smaller.

\subsection{Axial linear power and peaking factors}\label{sec:res-power}

\begin{figure}[htbp]
  \centering
  \includegraphics[width=0.65\textwidth]{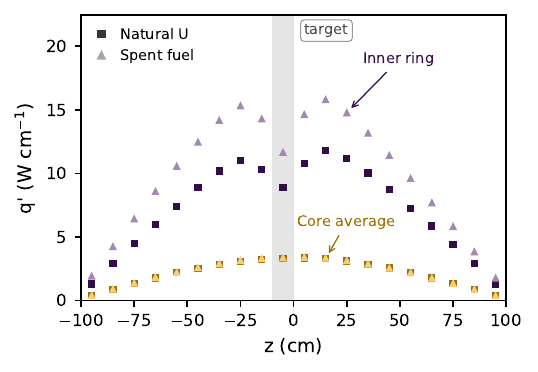}
  \caption{Linear power $q'$ in 10~cm axial cells for the inner ring of pins next to the cavity and for the average over all fuel pins, for natural uranium (dark squares) and spent fuel (light triangles), driven by one 400~kW S-BEND beamline. Each point is the average over the pins of the group; error bars, one standard deviation (statistical). The grey band marks the position of the converter.}
  \label{fig:linpower}
\end{figure}

Figure~\ref{fig:linpower} shows the linear power $q'$ of the inner ring of pins, next to the cavity, where the flux is highest, and its average over all fuel pins. The core-averaged linear power is $\bar{q}'=2.2$~W~cm$^{-1}$ in both designs, with an axial profile peaking at about 3.4~W~cm$^{-1}$ near mid-length. The inner ring operates at three to four and a half times this value. Its axial profile shows two maxima, upstream and downstream of the converter, separated by a local minimum at the position of the converter, where the uranium block replaces the heavy water of the cavity next to the inner ring. The higher maximum, at $z=15$~cm, defines the peak linear power $q'_\mathrm{max}=11.8\pm0.2$~W~cm$^{-1}$ for natural uranium and $15.8\pm0.2$~W~cm$^{-1}$ for spent fuel, giving power peaking factors $F_q=5.4$ and 7.2. 

%% file: Tables/key_results.tex
\begin{table}[htbp]
\centering
\caption{The two synchrotron-driven subcritical cores: natural UO$_2$ and PWR spent fuel (30~GWd/t, 5~years cooling, OECD/NEA Phase~I-A composition). Each lattice pitch gives $k_\mathrm{eff} = 0.985$; the photon beam (S-BEND spectrum, 1--100~MeV) and the converter target are the same. Uncertainties are 1$\sigma$ statistical. Photonuclear data from IAEA/PD-2019. $Y_1$: first-generation photoneutrons per source photon, from the beam and its shower (run without fission neutrons); $Y_n$: all photoneutrons per source photon at steady state, including those released by fission and capture gammas. $P$: fissions of each nuclide times its recoverable energy per fission ($^{235}$U 202.4, $^{238}$U 206.0, $^{239}$Pu 211.1, $^{241}$Pu 214.3~MeV), photofission included. $q'_\mathrm{max}$: peak of the axial linear power averaged over the inner ring of pins (next to the cavity), in 10~cm cells; $F_q = q'_\mathrm{max}/\bar{q}'$.}
\label{tab:key_results}
\small
\begin{tabularx}{\linewidth}{Xcc}
\toprule
Quantity & Natural U & Spent fuel \\
\midrule
\multicolumn{3}{l}{\textit{Core}} \\
Lattice pitch $p$ (cm) & $2.835$ & $2.450$ \\
Heavy metal $m_\mathrm{HM}$ (t) & $9.42$ & $9.25$ \\
Heavy water $m_\mathrm{D_2O}$ (t) & $16.1$ & $11.6$ \\
\midrule
\multicolumn{3}{l}{\textit{Photon beam and converter}} \\
Beam power $P_\mathrm{beam}$ (kW) & \multicolumn{2}{c}{$400$} \\
Photon intensity $S_0$ (s$^{-1}$) & \multicolumn{2}{c}{$3.76\times10^{17}$} \\
First-generation photoneutron yield $Y_1$ & $(6.645 \pm 0.020)\times10^{-3}$ & $(6.65 \pm 0.02)\times10^{-3}$ \\
External neutron source $S = S_0 Y_1$ (s$^{-1}$) & $(2.498 \pm 0.007)\times10^{15}$ & $(2.498 \pm 0.008)\times10^{15}$ \\
\midrule
\multicolumn{3}{l}{\textit{Subcritical core}} \\
Multiplication factor $k_\mathrm{eff}$ & $0.98491 \pm 0.00020$ & $0.9850 \pm 0.0002$ \\
Steady-state photoneutron yield $Y_n$ & $(7.39 \pm 0.02)\times10^{-3}$ & $(7.46 \pm 0.03)\times10^{-3}$ \\
Source multiplication factor $k_s$ & $0.99246 \pm 0.00010$ & $0.99301 \pm 0.00011$ \\
Source importance $\varphi^*$ & $1.90 \pm 0.04$ & $2.04 \pm 0.05$ \\
\midrule
\multicolumn{3}{l}{\textit{Power}} \\
Fission power $P$ (MW) & $4.37 \pm 0.06$ & $4.41 \pm 0.07$ \\
Energy gain $G = P/P_\mathrm{beam}$ & $10.92 \pm 0.14$ & $11.03 \pm 0.16$ \\
Peak linear power $q'_\mathrm{max}$ (W cm$^{-1}$) & $11.79 \pm 0.16$ & $15.8 \pm 0.2$ \\
Power peaking factor $F_q$ & $5.4$ & $7.2$ \\
\bottomrule
\end{tabularx}
\end{table}

%% file: Sections/03_discussion.tex
\section{Discussion}\label{sec:discussion}

The two cores reach the same gain with very different fissile inventories. The natural-uranium core holds 67~kg of $^{235}$U in 9.42~t of heavy metal, whereas the spent-fuel core holds 116~kg of $^{235}$U, 56~kg of $^{239}$Pu and 10~kg of $^{241}$Pu in 9.25~t, 182~kg of fissile nuclides in all. In a source-driven core the fissile inventory does not set the gain, which depends mainly on the subcriticality and on the importance of the source (Eq.~\eqref{eqn:power}). It sets instead the lattice in which the subcriticality is reached, which is wider for natural uranium. One design, adapted in its pitch alone, therefore accepts fuel from either end of the fuel cycle, and neither fuel passes through uranium enrichment or plutonium separation, the sensitive elements of a nuclear fuel cycle~\cite{IAEA_INFCIRC640}.

The spent-fuel core does not, however, eliminate waste. Minor actinides, key contributors to the long-term radiotoxicity of spent fuel~\cite{Cammi_2025}, are transmuted effectively only by fission, for which the neutron economy of a fast spectrum is better suited~\cite{IAEA_TECDOC985}, as in MYRRHA~\cite{Baeten_2014}; the thermal cores proposed here burn the fissile isotopes of the spent fuel rather than its minor actinides. Their role in the back end of the fuel cycle is rather that of a buffer between discharge and disposal. Spent fuel already spends decades in interim storage: periods of 100~years and more are now considered, and some countries are preparing for up to 300~years~\cite{IAEA_TECDOC1293_2002, IAEA_PASSED}. In a subcritical core, part of this period produces energy, until the residual fissile content can no longer sustain a useful gain and the fuel returns, unreprocessed, to the route it would have followed anyway. The specific power of the present cores, 0.47~W per gram of heavy metal, is about one hundredth of that of a pressurised-water reactor, so that the fuel accumulates only about 0.17~GWd~t$^{-1}$ per year; the residence time itself requires burnup calculations that are beyond the scope of this work. A facility of 50 cores would hold about 460~t of spent fuel producing power.

The scale of the concept is set by the storage ring. The proposed ring, with an electron energy of 32~GeV and a length of about 5~km, accommodates up to 50 photon beamlines~\cite{SYNERGY_2026}, each fed by its own S-BEND, which can be switched on or off without disturbing the stored beam (Section~\ref{sec:m-source}). The beam power of 400~kW per beamline is that of the present ring design, a realistic value rather than an upper limit: at a given subcriticality the fission power scales with the beam power, so that a more intense beam, from a higher stored current, would drive a proportionally more powerful core, within the heat load that the converter and the RF system can sustain. One ring thus drives 50 cores like those of Table~\ref{tab:key_results}, which together produce about 220~MW of fission power from 20~MW of photons, and which can be started, stopped and refuelled individually and hold different fuels. The 50 cores of a ring would hold 580--800~t of heavy water, the inventory of one to two CANDU~6 units~\cite{CanduEnergy_EC6_2015}, which, however, produce about 2.1~GW of heat.

Whether such a facility produces net electricity depends on $G$ and on two efficiencies: that of the conversion of heat into electricity, $\eta_{th}$, and that of the accelerator, $\eta_{acc}$, the ratio of the beam power to the electric power drawn from the grid. A fraction $f=1/(G\,\eta_{th}\,\eta_{acc})$ of the electricity generated returns to the accelerator, and the system produces more energy than the accelerator consumes when $f<1$~\cite{IAEA_TECDOC1766}. An electron-driven thorium molten-salt core was found slightly short of this condition, even for a lossless accelerator~\cite{Rummana_2024}. In a storage ring the radio-frequency (RF) system only restores the energy radiated by the circulating electrons~\cite{SYNERGY_2026}: in the SLS~2.0 light source the radiated power reaches about 46\% of the electric power drawn by the RF chain~\cite{Braun:2021sls2tdr}, which gives $\eta_{acc}\approx0.38$ per unit of beam power for the present ring, where 400 of the 481~kW radiated in each arc reach the core. With $G=11.0$ and $\eta_{th}=0.32$, the gross efficiency of a CANDU~6 heavy-water power reactor~\cite{WNA_Cernavoda}, $f\approx0.75$: a facility with 50 beamlines would generate about 70~MW of electricity and return about 53~MW to the RF system of the ring, before the magnets, the cryogenics and the auxiliary systems are counted. This efficiency requires a coolant at about 300 °C; in the present lattice, in which the heavy water is also the moderator, a hot pressurised coolant would have to be separated from a cold moderator, as in a CANDU. Klystrons with efficiencies of up to 80\%, under development for future circular colliders~\cite{Nisa_2023}, would raise $\eta_{acc}$ and lower $f$ further, towards the value of about 0.37 of a proton-driven design of 800~MW of thermal power~\cite{Dorda_2025}.

Table~\ref{tab:ads} compares the synchrotron-driven cores with the two accelerator-driven systems closest to them: the neutron source of KIPT, a thermal core driven by the photoneutrons of an electron linac~\cite{zelinsky2013nsc, Gohar_2022}, and MYRRHA, a fast core driven by spallation~\cite{Baeten_2014}. Both bring a charged-particle beam in vacuum to a target at the centre of the core. In MYRRHA the vacuum chambers of the beam line enter the core~\cite{Saugnac_2011}, at the centre of which the spallation target assembly receives the beam~\cite{Baeten_2014}. There, a window that receives up to 2.4~MW on an 85~mm footprint~\cite{Saugnac_2011} separates the vacuum of the beam tube from the lead--bismuth coolant, which also serves as spallation target~\cite{Keijers_2015}. Irradiation, lead--bismuth corrosion and thermal fatigue limit the life of the window, which accumulates up to 22~dpa in a 90-day cycle~\cite{Keijers_2015}. At KIPT the electrons reach the target, at the centre of the core~\cite{Karnaukhov_2024}, through a vacuum channel closed by a water-cooled aluminium-alloy window, and the target, of uranium or tungsten plates cooled by water, is designed to remove the whole 100~kW of the beam~\cite{zelinsky2013nsc}. A photon beam needs neither: photons of 1--20~MeV lose less than 1\% of their intensity per metre of air~\cite{Hubbell_2004}, so that the beamline can be closed by a window placed outside the core, away from its neutron flux, and the channel inside the core need not be evacuated. The converter, which receives 370~kW in about 3~L of uranium, would be built like the targets of electron-driven sources, of thin clad plates cooled by a forced flow of heavy water; its thermal design is beyond the scope of this work.

The three systems also differ in the intensity of their source per unit of beam power. Spallation releases 14.6 neutrons per 600~MeV proton~\cite{Keijers_2015}, about $1.5\times10^{14}$~n~s$^{-1}$ per kW, so that the MYRRHA beam of at most 2.4~MW drives up to 100~MW~\cite{Baeten_2014, Bouly_2014}, a gain of about 40, even 5000~pcm below criticality. Photonuclear sources are twenty to fifty times weaker for the same beam power~\cite{IAEA_TECDOC1766}: the S-BEND photons release $6.2\times10^{12}$~n~s$^{-1}$ per kW of beam power in the converter, about twice the $3.0\times10^{12}$ of the uranium target of KIPT~\cite{zelinsky2013nsc}. In an electron-driven source the electrons stop in the converter, where they lose part of their energy by ionisation and radiate a bremsstrahlung spectrum that extends from zero to the full beam energy, so that only part of the beam power reaches the giant dipole resonance; the S-BEND spectrum, with $E_c=10$~MeV, is instead chosen to maximise the photons between 10 and 20~MeV (Section~\ref{sec:m-source}), and the electrons that emit it are not lost but stay in the ring. Each synchrotron-driven core thus reaches a gain of about 11, against about 2 at KIPT, whose 100~kW beam releases 192~kW in its core~\cite{zelinsky2013nsc}, while remaining below the gain of MYRRHA, whose more intense spallation source outweighs its deeper subcriticality.

The synchrotron-driven cores operate 1500~pcm below criticality, KIPT at least 2000~pcm and MYRRHA 5000~pcm. The subcriticality of an accelerator-driven core is set by the largest positive insertion of reactivity that it must accommodate without becoming critical~\cite{Cahalan_2002}. MYRRHA was designed at $k_\text{eff}=0.95$, the level accepted for the storage of fuel, so as to remain subcritical after any such insertion~\cite{IAEA_TECDOC1766}. In a fast core a compaction of the fuel can add reactivity~\cite{Petrovic_2023}, and the large minor-actinide loadings envisaged for transmutation reduce the delayed-neutron fraction and the Doppler effect and increase the void effect~\cite{Granget_2006}. The thermal cores proposed here are less exposed to these effects. More than 97\% of their heavy metal is $^{238}$U, a fertile nuclide that provides a negative Doppler feedback~\cite{Cahalan_2002}, and the heavy water is at the same time the coolant and the moderator of an under-moderated lattice, in which a decrease in its density, by heating, boiling or loss of coolant, reduces $k_\text{eff}$. The margin of KIPT is of the same order as ours: its subcritical assembly was designed to remain below $k_\text{eff}=0.98$ in normal, abnormal and accident conditions~\cite{Karnaukhov_2024}, and was licensed for its physical start-up in 2020~\cite{Gohar_2022}.

Finally, the heavy-water cores admit natural uranium and spent fuel, whereas KIPT uses uranium enriched to 19.7\%~\cite{zelinsky2013nsc} and MYRRHA mixed-oxide fuel from reprocessed plutonium, in a core intended also to demonstrate waste transmutation~\cite{Baeten_2014}: the synchrotron-driven cores complement rather than replace these systems.

\input{Tables/ads_comparison}

The photon-driven cores of Sun et al.~\cite{Sun_2022} and Hu et al.~\cite{Hu_2025} share the use of a photon beam to drive a photoneutron source, but aim at transmutation. In the Gamma Factory design, 16.4~MW of photons drive a single 500~MW core, a gain of about 30~\cite{Hu_2025}, against 11 here. Neither source exists yet: the Gamma Factory needs a fourfold increase of the radio-frequency power of the LHC~\cite{Hu_2025}, and the storage ring needs S-BENDs at the limit of present dipoles. The two approaches are complementary: an intense, narrow-band beam suits one large core for transmutation, whereas the 50 beamlines of a ring drive 50 small cores that produce energy from unenriched fuel.

The source depends on the photonuclear data, whereas the simplifications of the model overestimate the multiplication of the core. The results above use the photonuclear data of IAEA/PD-2019. With those of the ENDF7u library, whose evaluations end at 20~MeV for $^{238}$U and at 30~MeV for $^2$H (Section~\ref{sec:m-model}), the external source $S$ of both cores is 12\% lower and the gains are $G=9.65\pm0.17$ for natural uranium and $9.58\pm0.15$ for spent fuel, a conservative estimate. The cores are modelled cold and clean, with pure heavy water and without temperature feedback, xenon or samarium. Since the lattice is under-moderated, the cold state is more reactive than the hot one, and the subcriticality of $k_\text{eff}=0.985$ must be ensured in it; at operating temperature and with xenon and samarium at equilibrium, $k_\text{eff}$ and $G$ would be lower. 

In conclusion, a 400~kW beam of synchrotron photons drives a subcritical heavy-water core of natural uranium or of spent fuel to about 4.4~MW of fission power, a gain of about 11, 1500~pcm below criticality. The same lattice, adapted in its pitch alone, accepts fuel that has passed neither through enrichment nor through reprocessing. A storage ring with 50 beamlines would drive 50 such cores, with about 220~MW of fission power and no vacuum or beam window inside them, and would generate about 70~MW of electricity against the 53~MW drawn by its RF system. Within the limits of the present model, synchrotron radiation is thus a candidate driver for thermal subcritical cores, complementary to the proton-driven fast systems designed for transmutation.

%% file: Tables/ads_comparison.tex
\begin{table}[htbp]
\centering
\caption{Synchrotron-driven cores: this work, with a storage ring of 50 beamlines~\cite{SYNERGY_2026}; heat in the target, part of the 384~kW carried by the photons above 1~MeV that is deposited in the converter. KIPT: design values of the neutron source facility with the uranium target~\cite{zelinsky2013nsc, Karnaukhov_2024}; heat in the target, the beam power that the target cooling is designed to remove, and fission power, the energy release of the subcritical assembly~\cite{zelinsky2013nsc}. MYRRHA: design values at the maximum current of 4~mA~\cite{Baeten_2014, Bouly_2014, Saugnac_2011, Keijers_2015}; heat in the target, 62\% of the beam power, and source, 14.6 neutrons per proton~\cite{Keijers_2015}.}
\label{tab:ads}
\small
\setlength{\tabcolsep}{4pt}
\begin{tabularx}{\linewidth}{@{}l>{\centering\arraybackslash}X>{\centering\arraybackslash}X>{\centering\arraybackslash}X@{}}
\toprule
 & Synchrotron-driven (this work) & KIPT & MYRRHA \\
\midrule
Accelerator & 32~GeV electron ring & 100~MeV electron linac & 600~MeV proton linac \\
Cores per accelerator & 50 & 1 & 1 \\
Vacuum in the core & no & yes & yes \\
Beam window in the core & none & Al alloy, water-cooled & 22~dpa per 90~days \\
Target & uranium & uranium or tungsten & lead--bismuth \\
Spectrum, coolant & thermal, D$_2$O & thermal, H$_2$O & fast, lead--bismuth \\
Fuel & natural UO$_2$ or spent fuel & UO$_2$, 19.7\% $^{235}$U & MOX \\
$k_\text{eff}$ & 0.985 & $\le0.98$ & 0.95 \\
Purpose & power, spent-fuel buffer & research & research, transmutation \\
\midrule
\multicolumn{4}{@{}l}{\textit{Per core}} \\
Beam power (MW) & 0.4 & 0.1 & 2.4 \\
Heat in the target (MW) & 0.37 & 0.1 & 1.5 \\
Source (n~s$^{-1}$) & $2.5\times10^{15}$ & $3.0\times10^{14}$ & $3.6\times10^{17}$ \\
Fission power (MW) & 4.4 & 0.19 & 100 \\
\midrule
\multicolumn{4}{@{}l}{\textit{Per accelerator}} \\
Beam power (MW) & 20 & 0.1 & 2.4 \\
Fission power (MW) & 220 & 0.19 & 100 \\
\bottomrule
\end{tabularx}
\end{table}

%% file: Sections/04_methods.tex
\section{Methods}\label{sec:methods}

\subsection{Synchrotron photon source}\label{sec:m-source}

The photon source considered in this work is based on high-field superconducting bending magnets (S-BENDs) installed in a 32~GeV electron storage ring operating at a beam current of 26.7~mA~\cite{SYNERGY_2026}. Each S-BEND, with a magnetic field of 14.7 T and an effective length of 5.37 cm, generates 400 kW of synchrotron radiation with a critical photon energy of 10 MeV.

The proposed storage ring has a circumference of approximately 5.47 km and consists of 50 arcs, each equipped with an S-BEND photon source. Synchrotron radiation losses are compensated by superconducting RF cavities. Electron storage rings operating at comparable beam energies and with similar circumferences have already been successfully implemented in particle colliders, such as the 6.3 km HERA electron ring, which operated at 27.5~GeV~\cite{Willeke2021}. A detailed description of the proposed storage ring, including the magnetic lattice, S-BEND configuration, beam dynamics, and RF system, is beyond the scope of the present paper and will be addressed in a dedicated publication.

\begin{figure}[htbp]
  \centering
  \includegraphics[width=0.7\textwidth]{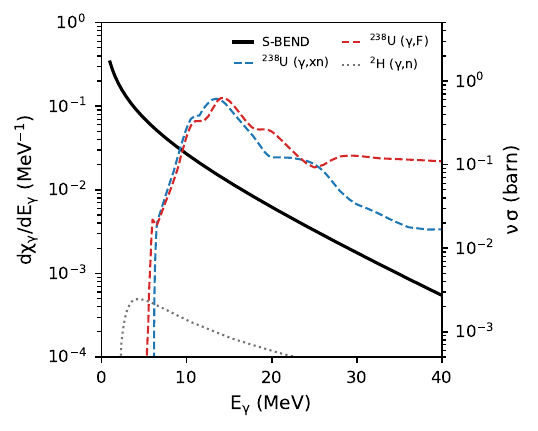}
  \caption{Left axis: normalised spectrum $\mathrm{d}\chi_\gamma/\mathrm{d}E_\gamma$ of the S-BEND photons for $E_c=10$~MeV (Eq.~\eqref{eqn:sr_spectrum}, one photon between 1 and 100~MeV; black line). Right axis: neutron production cross sections $\nu\sigma$, the cross section times the mean number of neutrons emitted, of the $^{238}$U$(\gamma,xn)$ and $^{238}$U$(\gamma,F)$ reactions (dashed lines) and of $^{2}$H$(\gamma,n)$ (dotted line), from the IAEA/PD-2019 library. The $(\gamma,xn)$ reaction groups the non-fission channels, whose neutron multiplicity is zero below 6.15~MeV and rises to about two above the $(\gamma,2n)$ threshold of 11.3~MeV. Shown up to 40~MeV, below which lie 99.5\% of the photons and 96\% of the power of the source.}
  \label{fig:source_xs}
\end{figure}

For the reactor simulations, the photon energy distribution is described by the standard synchrotron radiation spectrum~\cite{Wiedemann_2015}, with a critical energy of 10~MeV. Between two photon energies $E_1$ and $E_2$, the emitted photons have the normalised spectrum
\begin{equation}
\label{eqn:sr_spectrum}
\frac{\mathrm{d}\chi_\gamma}{\mathrm{d}E_\gamma} = \frac{P_{rad}}{S_0\,E_c\,E_\gamma}\;\mathcal{S}\!\left(\frac{E_\gamma}{E_c}\right),
\qquad
\mathcal{S}(y) = \frac{9\sqrt{3}}{8\pi}\,y\int_y^{\infty} K_{5/3}(x)\,\mathrm{d}x,
\end{equation}
where $P_{rad}$ is the radiated power, $E_c$ the critical energy, $K_{5/3}$ the modified Bessel function of the second kind, $\mathcal{S}$ the universal function of synchrotron radiation, normalised as $\int_0^\infty\mathcal{S}(y)\,\mathrm{d}y=1$, and $S_0$ the photon emission rate between $E_1$ and $E_2$, which normalises the spectrum to one photon in this range. The resulting photon spectrum, together with the relevant uranium photonuclear cross sections, is shown in Fig.~\ref{fig:source_xs}. In particular, the 10–20 MeV photon energy range, where neutron production cross sections are highest, is efficiently covered by the proposed source. Photons are sampled in the energy range 1–100 MeV, which accounts for approximately 384 kW of the total 400 kW emitted by each S-BEND. This corresponds to a photon emission rate of approximately $S_0=3.76\times10^{17}$~photons~s$^{-1}$, with a mean photon energy of 6.4~MeV. At the reactor entrance, the photon beam is assumed to be uniformly distributed over a rectangular area of $50\times2$ cm$^2$ and directed along the beam channel axis.

\subsection{Photonuclear neutron production}\label{sec:m-photonuclear}

Photons interact with the atomic electrons of the converter and with its nuclei. From the neutron separation energy to about 30~MeV, nuclear absorption is dominated by the giant dipole resonance, a collective oscillation of the protons against the neutrons~\cite{Bergere1977}, which in $^{238}$U, a deformed nucleus, splits into two peaks near 11 and 14~MeV, while total absorption is dominated by pair production for high-Z materials. The excited nucleus decays by emitting one or more neutrons above 6.15~MeV, the $(\gamma,xn)$ reaction, with a mean multiplicity that rises from one to about two above the $(\gamma,2n)$ threshold of 11.3~MeV, or by photofission, which releases on average 3.5 neutrons per fission in the converter. In heavy water, deuterium breaks up through D$(\gamma,n)$ above 2.224~MeV. We call photoneutrons the neutrons released by all these reactions, whose neutron production cross sections $\nu\sigma$ are shown in Fig.~\ref{fig:source_xs}.

Even at the resonance peak, the photonuclear cross section of uranium is a few per cent of the electromagnetic one, which pair production dominates. A beam photon therefore most often converts into an electron--positron pair within about one radiation length, $X_0=0.32$~cm in uranium, and the pair radiates bremsstrahlung photons that convert in turn: the beam energy is shared among the particles of an electromagnetic shower, whose photons fall progressively below the photonuclear thresholds, and a photoneutron can be released at every step of the cascade. For a converter that absorbs the whole shower, the yield from the first collision (fc) of the beam photons is
\begin{equation}
\label{eqn:photoneutron_yield}
Y_\mathrm{fc} = \int_{E_1}^{E_2} \frac{\mathrm{d}\chi_\gamma}{\mathrm{d}E_\gamma}\,\sum_r \nu_r(E_\gamma)\,\frac{\Sigma_r(E_\gamma)}{\Sigma_t(E_\gamma)}\,\mathrm{d}E_\gamma,
\end{equation}
where $\mathrm{d}\chi_\gamma/\mathrm{d}E_\gamma$ is the normalised photon spectrum of Eq.~\eqref{eqn:sr_spectrum}, $\Sigma_r$ and $\nu_r$ the macroscopic cross section and the neutron multiplicity of each photonuclear reaction $r$ (the mean number of neutrons emitted, for $(\gamma,xn)$ as for photofission), and $\Sigma_t$ the total macroscopic cross section, atomic and nuclear~\cite{Loi_2026}. The secondary photons of the shower add to $Y_\mathrm{fc}$. Following the whole shower, the transport calculation gives $Y_1=6.65\times10^{-3}$ photoneutrons per source photon (Table~\ref{tab:key_results}), half of them from photofission (Supplementary Table~S1). We call $Y_1$ the first-generation yield, since it does not include the photoneutrons released by the gammas of the core (Section~\ref{sec:m-feedback}), and $S=S_0Y_1=2.50\times10^{15}$~neutrons~s$^{-1}$ is the external neutron source of the core. The converter, 10~cm or 32 radiation lengths thick, absorbs about 370~kW of the 384~kW of photons that enter it.

\subsection{Subcritical multiplication and power}\label{sec:m-subcritical}

Without an external source, the neutron population of a multiplying medium converges to the fundamental mode $\phi$, solution of
\begin{equation}
\label{eqn:eigen}
\mathcal{L}\phi = \frac{1}{k_\text{eff}}\,\mathcal{F}\phi
\end{equation}
with the largest eigenvalue $k_\text{eff}$, where $\mathcal{F}$ is the fission production operator and $\mathcal{L}$ accounts for all other processes: transport, scattering including $(n,xn)$ reactions, absorption and leakage. A core with $k_\text{eff}<1$ driven by an external neutron source $Q$ reaches the steady state
\begin{equation}
\label{eqn:driven}
\mathcal{L}\phi_s = \mathcal{F}\phi_s + Q,
\end{equation}
which we characterise by the source multiplication factor~\cite{Gandini_2002,Seltborg_2003}
\begin{equation}
\label{eqn:ks}
k_s = \frac{F_n}{F_n+S},
\end{equation}
the fraction of fission neutrons among all the neutrons produced in the core, where $F_n=\langle\mathcal{F}\phi_s\rangle$ is the production rate of fission neutrons and $S=\langle Q\rangle$ the intensity of the source. Each source neutron leads to $M=1/(1-k_s)$ neutrons in total, itself included.

Unlike $k_\text{eff}$, $k_s$ depends on where and with which energy the source neutrons are emitted. Multiplying Eq.~\eqref{eqn:driven} by the adjoint fundamental mode $\phi^\dagger$, solution of $\mathcal{L}^\dagger\phi^\dagger=\mathcal{F}^\dagger\phi^\dagger/k_\text{eff}$, and integrating over phase space gives $\langle\phi^\dagger Q\rangle=(1/k_\text{eff}-1)\,\langle\phi^\dagger\mathcal{F}\phi_s\rangle$, so that
\begin{equation}
\label{eqn:phistar}
\varphi^* \equiv \frac{\langle\phi^\dagger Q\rangle/S}{\langle\phi^\dagger\mathcal{F}\phi_s\rangle/F_n} = \frac{1/k_\text{eff}-1}{1/k_s-1}.
\end{equation}
The source importance $\varphi^*$ is the mean importance of the source neutrons relative to that of the fission neutrons of the driven core, called neutron worth in refs.~\cite{Sun_2022, Hu_2025}. It exceeds one when the source neutrons are emitted where the importance is higher than on average, as at the centre of the core, and each of them then generates $\varphi^*$ times as many fissions as an average fission neutron. For a bare homogeneous core in one-group diffusion theory, with a point source at its centre and the fission neutrons distributed as the fundamental mode, Eq.~\eqref{eqn:phistar} gives $\varphi^*=(4/\pi)^3\approx2.06$ whatever the size of the core, close to the values found for the two cores (Table~\ref{tab:key_results}).

The fission power is
\begin{equation}
\label{eqn:pcore}
P_{core} = \sum_i \epsilon_i f_i,
\end{equation}
where $f_i$ is the rate of neutron-induced fissions and photofissions of nuclide $i$ in the whole core and $\epsilon_i$ its recoverable energy per fission: 202.4, 206.0, 211.1 and 214.3~MeV for $^{235}$U, $^{238}$U, $^{239}$Pu and $^{241}$Pu~\cite{Ma_2013}. These values include the decay energy of the fission products at equilibrium and the energy released by neutron captures, and exclude the antineutrinos. The $^{239}$Pu value is used for the other isotopes of plutonium, neptunium and americium, and the $^{238}$U value for $^{234}$U and $^{236}$U, which together account for less than 1\% of the fissions. The energy that the photon beam deposits in the core, about 370~kW, is not part of $P_{core}$. Since fission neutrons are produced at the rate $F_n=S\,\varphi^*k_\text{eff}/(1-k_\text{eff})$ (Eqs.~\eqref{eqn:ks} and \eqref{eqn:phistar}), the power can be written as
\begin{equation}
\label{eqn:power}
P_{core} \simeq S\,\frac{\bar\epsilon}{\bar\nu}\,\varphi^*\,\frac{k_\text{eff}}{1-k_\text{eff}},
\end{equation}
with $\bar\nu$ the mean number of neutrons and $\bar\epsilon$ the mean recoverable energy per fission; photofission, 0.3\% of the power, is neglected. For $\varphi^*=1$, Eq.~\eqref{eqn:power} reduces to the point-reactor model, in which the power is set by the subcriticality alone.

The linear power $q'$ of the pins is obtained from pin-by-pin fission rates in axial cells of 10~cm, multiplied by $\bar\epsilon$. Its peak value $q'_\mathrm{max}$ is the maximum of the axial profile averaged over the inner ring of pins, adjacent to the cavity, and the peaking factor is $F_q=q'_\mathrm{max}/\bar{q}'$, with $\bar{q}'$ the average over all fuel pins.

\subsection{Photoneutrons from the gammas of the core}\label{sec:m-feedback}

The gammas emitted in fission and radiative capture also release photoneutrons, mainly through D$(\gamma,n)$ in the heavy water. Each fission emits about seven prompt gammas, 0.55 of which on average lie above the 2.224~MeV threshold; a capture in $^{238}$U adds 0.69 such gammas, and a capture in deuterium one gamma of 6.25~MeV. These photoneutrons are produced by the neutron population of the core and belong to the multiplication, not to the external source. We separate the two contributions with a second fixed-source calculation in which neutron-induced fission is switched off and photofission is kept: its photoneutrons, released by the beam photons and their shower, give the first-generation yield $Y_1$ and the external source $S=S_0Y_1$. In the full calculation, the steady-state yield $Y_n$ also includes the photoneutrons released by the gammas of the core, $Y_\gamma=Y_n-Y_1$ per source photon, 11--12\% of $Y_1$ (Supplementary Table~S1). With their production rate $S_0Y_\gamma$, the source multiplication factor then becomes
\begin{equation}
\label{eqn:ks_feedback}
k_s = \frac{F_n+S_0Y_\gamma}{F_n+S_0Y_\gamma+S}.
\end{equation}
The eigenvalue calculation transports neutrons only and misses this contribution. We correct its multiplication factor as
\begin{equation}
\label{eqn:kprime}
k' = k_\text{eff}\left(1+\frac{S_0Y_\gamma}{F_n}\right),
\end{equation}
assuming that the photoneutrons released per fission neutron are the same in the fundamental mode as in the driven core, and use $k'$ in place of $k_\text{eff}$ in Eqs.~\eqref{eqn:phistar} and \eqref{eqn:power}. The correction is about 85~pcm in both cores. The evaluated data contain the prompt fission gammas only, so that the photoneutrons released by the delayed gammas of the fission products are not included.

\subsection{Fuel compositions}\label{sec:m-fuel}

The natural-uranium core is fuelled with UO$_2$ at 10.2~g~cm$^{-3}$, with the natural isotopic composition of uranium. The spent fuel is that of Phase~I-A of the OECD/NEA Burnup Credit Criticality Benchmark~\cite{Takano_1994}: fuel from a Westinghouse $17\times17$ PWR assembly with an initial enrichment of 3.6~wt\% $^{235}$U, discharged at 30~GWd~t$^{-1}$ and cooled for 5~years. The benchmark gives the atom densities of 12 actinides, oxygen and the 15 fission products most important for burnup credit (Supplementary Table~S2), for a density of 10.1~g~cm$^{-3}$; the other fission products are not included. The fuel is loaded with this composition, without reprocessing. The spent-fuel pins of the model are not actual PWR rods: they have the fuel and cladding radii of a $17\times17$ rod and the composition of the benchmark, but the 200~cm length of the model, whereas the active length of a PWR rod is 366~cm. The model therefore represents the fissile inventory of the spent fuel in a simplified geometry; a core of full-length rods would reach $k_\text{eff}=0.985$ with a tighter pitch. It contains 1.27~wt\% of $^{235}$U in uranium and 0.95~wt\% of plutonium in heavy metal, and its fissile content, $^{235}$U, $^{239}$Pu and $^{241}$Pu, is 1.96~wt\% of the heavy metal, against 0.71~wt\% for natural uranium.

\subsection{Monte Carlo model and nuclear data}\label{sec:m-model}

All calculations use OpenMC~\cite{Romano_2015} with photonuclear physics, from the development branch \texttt{photonuclear-physics}~\cite{Stein_2024} (version 0.16.1-dev138, commit 71433a7), in which photonuclear reactions release neutrons during the photon transport. Its photoneutron production agrees within 0.7\% with that of MCNPX when the same photonuclear data are used, and has also been compared with FLUKA and with a first-collision analytical solution~\cite{Loi_2026}. Electrons and positrons are not transported. Their bremsstrahlung is treated with the thick-target approximation~\cite{Lund_2018}: each charged particle is assumed to slow down in the material in which it is created, where it emits, along its direction of flight, the bremsstrahlung photons that it would produce over its range and deposits the rest of its energy; positrons then annihilate into two photons of 0.511~MeV. Photons, electrons and positrons are followed down to 1~MeV: below this energy a photon cannot release a neutron, and a charged particle cannot emit a photon that can, the lowest photonuclear threshold being 2.224~MeV, for deuterium.

Neutron and photon interaction data are taken from ENDF/B-VII.1~\cite{ENDFVII1}, with the thermal scattering laws of deuterium in heavy water and of graphite, and photonuclear data from the IAEA Photonuclear Data Library 2019 (IAEA/PD-2019)~\cite{Kawano_2020}, whose evaluations of $^{238}$U and $^2$H extend to 200 and 140~MeV, beyond the upper end of the source spectrum. The fixed-source calculations were repeated with the photonuclear data of the ENDF7u library~\cite{ENDF7U}, whose evaluations end at 20~MeV for $^{238}$U and at 30~MeV for $^2$H, with zero cross sections above these energies (Section~\ref{sec:discussion}). All materials are at 294~K: at the linear powers of the two cores, below 20~W~cm$^{-1}$, the fuel would be only a few tens of kelvin hotter than the coolant.

The core is a square lattice of $100\times100$ pin positions, 200~cm long, with a pitch of 2.835~cm (natural uranium) or 2.450~cm (spent fuel). The fuel has a radius of 0.41~cm and the cladding, natural zirconium at 6.5~g~cm$^{-3}$, an outer radius of 0.475~cm; coolant and moderator are pure D$_2$O at 1.1~g~cm$^{-3}$. A cavity of $21\times4$ (natural uranium) or $24\times4$ (spent fuel) pin positions at the centre of the lattice is filled with heavy water and contains the converter, natural uranium metal at 19.0~g~cm$^{-3}$, which occupies $19\times2$ or $22\times2$ positions ($53.9\times5.7$ or $53.9\times4.9$~cm$^2$) between $z=-10$ and 0~cm. Between the converter and the face of the core at $z=100$~cm, a $50\times2$~cm$^2$ channel, void in the model, brings the photon beam into the core. The lattice is surrounded by 1~cm of graphite, with vacuum boundary conditions on the outer surface. The two cores contain 9916 and 9904 fuel pins.

Three calculations are made for each core. An eigenvalue calculation with neutron transport only gives $k_\text{eff}$: 350 generations of 75\,000 neutrons, the first 75 of which are discarded, starting from a uniform source in the fuel; the Shannon entropy of the fission source is stable after about 25 generations. A fixed-source calculation with the photon beam as source gives the steady-state photoneutron yield $Y_n$ by reaction and region, the fission rates by nuclide and by pin, the neutron flux and spectra, and the energy deposition. A second fixed-source calculation without neutron-induced fission gives the first-generation yield $Y_1$ and the source $S$ (Section~\ref{sec:m-feedback}). Each fixed-source calculation follows $4\times10^7$ photons in 400 batches.

\subsection{Normalisation and statistics}\label{sec:m-norm}

The fixed-source tallies, scored per source photon, are multiplied by $S_0=3.76\times10^{17}$~s$^{-1}$ to obtain absolute rates, and the gain follows as $G=P_{core}/P_{beam}$ (Eq.~\eqref{eqn:gain}) with $P_{beam}=400$~kW, the total power radiated by the S-BEND. Uncertainties are one standard deviation of the mean over the batches. For $k_\text{eff}$, the larger of the value reported by OpenMC and that obtained from the means of blocks of 25 generations is taken, to account for the correlation between successive generations. Uncertainties are propagated to first order to $k_s$, $k'$, $\varphi^*$, $P_{core}$ and $G$, with the eigenvalue and fixed-source calculations treated as independent. All calculations use the same random-number seed, so that $Y_1$ and $Y_n$ are positively correlated and the uncertainty of $Y_\gamma$, obtained by adding theirs in quadrature, is conservative; for the same reason, the two cores, whose beam and converter are identical, have the same $Y_1$ within 0.03\%.

\subsection{Decomposition of the gain}\label{sec:m-gain}

Dividing Eq.~\eqref{eqn:power}, with $k'$ in place of $k_\text{eff}$ and $S=S_0Y_1$, by the beam power expresses the gain as the product of five factors,
\begin{equation}
\label{eqn:gain_decomposition}
G \simeq \frac{S_0}{P_{beam}}\cdot Y_1\cdot\varphi^*\cdot\frac{k'}{1-k'}\cdot\frac{\bar\epsilon}{\bar\nu}:
\end{equation}
the photons emitted per unit of beam energy, the first-generation photoneutrons per photon, the source importance, the multiplication of the fundamental mode and the energy released per fission neutron. The first two factors depend only on the beam and the converter and are the same in both cores, $9.40\times10^{11}$~J$^{-1}$ and $6.65\times10^{-3}$. The source importance is 7.4\% higher in the spent-fuel core, 2.04 against 1.90, and $k'/(1-k')$ is 0.5\% higher, 69.5 against 69.2, since both cores have the same $k_\text{eff}$ and nearly the same correction for the photoneutrons of their gammas. The energy released per fission neutron is 6.5\% lower, $207.2/2.677=77.4$~MeV against $202.4/2.444=82.8$~MeV, because the larger number of neutrons per fission of plutonium outweighs its larger recoverable energy. These differences compensate, and the gains, 11.03 and 10.92, differ by about 1\%.

%% file: Tables/photoneutrons.tex
\begin{table}[htbp]
\centering
\caption{Origin of the photoneutrons, per source photon. $Y_1$: first-generation photoneutrons, made by the beam and its shower (run without fission neutrons). $Y_n$: all photoneutrons at steady state in the driven core, including those released by fission and capture gammas. Channel values from $\sigma(E)\nu(E)$ tallies; totals from analog transport (the two agree within 1\%).}
\label{tab:photoneutrons}
\begin{tabular}{lrrrr}
\toprule
 & \multicolumn{2}{c}{First generation, $Y_1$} & \multicolumn{2}{c}{Steady state, $Y_n$} \\
\cmidrule(lr){2-3} \cmidrule(lr){4-5}
Channel & $10^{-3}$\,n/$\gamma$ & \% & $10^{-3}$\,n/$\gamma$ & \% \\
\multicolumn{5}{l}{\textit{Converter target}} \\
$^{238}$U($\gamma$,F) & $3.269$ & $48.8$ & $3.269$ & $43.9$ \\
$^{238}$U($\gamma$,xn) & $3.369$ & $50.3$ & $3.369$ & $45.3$ \\
\midrule
\multicolumn{5}{l}{\textit{Heavy water}} \\
$^{2}$H($\gamma$,n) & $0.005$ & $0.1$ & $0.722$ & $9.7$ \\
\midrule
Other sources & $0.058$ & $0.9$ & $0.084$ & $1.1$ \\
\midrule
Total & $6.645 \pm 0.020$ & $100$ & $7.39 \pm 0.02$ & $100$ \\
\bottomrule
\end{tabular}
\end{table}

%% file: Tables/sf_composition.tex
\begin{table}[htbp]
\centering
\caption{Atom densities of PWR spent fuel with an initial enrichment of 3.6~wt\% $^{235}$U, discharged at 30~GWd~t$^{-1}$ and cooled for 5~years, from Phase~I-A of the OECD/NEA Burnup Credit Criticality Benchmark~\cite{Takano_1994}.}
\label{tab:sf_composition}
\small
\begin{tabularx}{\linewidth}{l>{\centering\arraybackslash}Xl>{\centering\arraybackslash}X}
\toprule
\multicolumn{2}{l}{Actinides and oxygen} & \multicolumn{2}{l}{Fission products} \\
\cmidrule(lr){1-2} \cmidrule(lr){3-4}
Nuclide & $N$ (b$^{-1}$\,cm$^{-1}$) & Nuclide & $N$ (b$^{-1}$\,cm$^{-1}$) \\
\midrule
$^{234}$U & $4.9750\times10^{-6}$ & $^{95}$Mo & $4.1250\times10^{-5}$ \\
$^{235}$U & $2.8374\times10^{-4}$ & $^{99}$Tc & $4.0589\times10^{-5}$ \\
$^{236}$U & $9.9361\times10^{-5}$ & $^{101}$Ru & $3.8224\times10^{-5}$ \\
$^{238}$U & $2.1760\times10^{-2}$ & $^{103}$Rh & $2.4046\times10^{-5}$ \\
$^{237}$Np & $1.0281\times10^{-5}$ & $^{109}$Ag & $2.7336\times10^{-6}$ \\
$^{238}$Pu & $2.8060\times10^{-6}$ & $^{133}$Cs & $4.4238\times10^{-5}$ \\
$^{239}$Pu & $1.3521\times10^{-4}$ & $^{143}$Nd & $3.1281\times10^{-5}$ \\
$^{240}$Pu & $4.2930\times10^{-5}$ & $^{145}$Nd & $2.4370\times10^{-5}$ \\
$^{241}$Pu & $2.3243\times10^{-5}$ & $^{147}$Sm & $7.5335\times10^{-6}$ \\
$^{242}$Pu & $7.4129\times10^{-6}$ & $^{149}$Sm & $1.7886\times10^{-7}$ \\
$^{241}$Am & $6.8518\times10^{-6}$ & $^{150}$Sm & $1.0403\times10^{-5}$ \\
$^{243}$Am & $1.3650\times10^{-6}$ & $^{151}$Sm & $5.7749\times10^{-7}$ \\
$^{16}$O & $4.6215\times10^{-2}$ & $^{152}$Sm & $4.2635\times10^{-6}$ \\
 &  & $^{153}$Eu & $3.3831\times10^{-6}$ \\
Total & $6.8866\times10^{-2}$ & $^{155}$Gd & $2.2348\times10^{-7}$ \\
\bottomrule
\end{tabularx}
\end{table}